\documentclass[aps,pre,twocolumn,amsmath,amssymb,superscriptaddress,floatfix,eqsecnum,showpacs]{revtex4}
\usepackage{graphicx}
\usepackage{dcolumn}
\usepackage{bm}
\bibstyle{apsrev.bib}

\newcommand{\be}{\begin{equation}}
\newcommand{\ee}{\end{equation}}
\newcommand{\bea}{\begin{eqnarray}}
\newcommand{\eea}{\end{eqnarray}}

\begin{document}

\title{Nonequilibrium phase transition in a driven Potts model with friction}

\author{Ferenc Igl\'oi}%
 \email{igloi@szfki.hu}
 \affiliation{Research Institute for Solid State Physics and Optics, 
P.O. Box 49, H-1525 Budapest, Hungary}
 \affiliation{Institute of Theoretical Physics,
Szeged University, H-6720 Szeged, Hungary}
 \affiliation{Groupe de Physique Statistique, D\'epartement Physique de la Mati\`ere et 
des Mat\'eriaux,
Institut Jean Lamour, CNRS---Nancy Universit\'e---UPV Metz,
BP 70239, F-54506 Vand\oe uvre l\`es Nancy Cedex, France}
\author{Michel Pleimling}
 \email{Michel.Pleimling@vt.edu}
 \affiliation{Department of Physics, Virginia Polytechnic Institute and State University,
Blacksburg, Virginia 24061-0435, USA}
\author{Lo\"{\i}c Turban}
 \email{loic.turban@ijl.nancy-universite.fr}
 \affiliation{Groupe de Physique Statistique, D\'epartement Physique de la Mati\`ere 
et des Mat\'eriaux,
Institut Jean Lamour, CNRS---Nancy Universit\'e---UPV Metz,
BP 70239, F-54506 Vand\oe uvre l\`es Nancy Cedex, France}

\date{\today}

\begin{abstract}
We consider magnetic friction between two systems of $q$-state Potts spins which are moving along their boundaries
with a relative constant velocity $v$. Due to the interaction between the surface spins there is a permanent energy
flow and the system is in a steady state which is far from equilibrium. The problem is treated analytically
in the limit $v=\infty$ (in one dimension, as well as in two dimensions for large-$q$ values) and for $v$ and $q$ finite by Monte Carlo simulations in two dimensions.
Exotic nonequilibrium phase transitions take place, the properties of which depend on 
the type of phase transition in equilibrium. When this latter transition is of first order, a sequence of second- and
first-order nonequilibrium transitions can be observed when the interaction is varied. 
\end{abstract}


\maketitle

\section{Introduction}
\label{sec:1}
Friction is a basic problem in physics with important technical implications~\cite{popov}. Friction between moving bodies and thus energy
dissipation may have magnetic contributions. This type of phenomenon takes place,
for example, in magnetic force microscopy when, performing a measurement, a magnetic tip is moved over the surface
of a magnetic material~\cite{tip1,tip2,tip3}. Recently, magnetic friction has been modeled in a simple setup~\cite{kadau},
where two Ising models
are put in close contact at their surfaces and move with a constant relative velocity $v$. The spins in the
surface layers interact through a coupling $K_f$ and the relaxation process due to phononic and electronic degrees
of freedom in the material are taken into account via a heat bath, at a fixed temperature $T$, to which all the
spins are coupled. In this system magnetic friction takes place and there is a permanent energy dissipation from
the surface to the heat bath. The macroscopic motion of the bodies represents a permanent perturbation
driving the system to a steady state which is far from equilibrium. Properties of this nonequilibrium state have been investigated by Monte Carlo simulations and --- in the limit $v=\infty$ --- by analytical methods in
different geometries~\cite{kadau,hucht,hilhorst}. In the one-dimensional (1D) case there is an order-disorder transition at $v=\infty$, but the system stays disordered for any finite $v$ in the thermodynamic limit. Fluctuation effects introduced by a finite size at $v=\infty$ have been recently studied by Hilhorst~\cite{hilhorst}.  In the two-dimensional (2D) case, when the
moving systems are in contact at their 1D surfaces, the order-disorder transition persists for any $v>0$. In all
cases the nonequilibrium phase transition is found to be of the mean-field type.

The spin degrees of freedom, which are involved in the friction, can have different symmetries and/or 
different types of interactions. As a consequence the equilibrium phase transition of the system at $v=0$
can be different from that of the Ising model, thereby influencing the behavior of the system out of equilibrium, when $v>0$.
In particular, the properties of the nonequilibrium phase transition could differ from those of a mean-field transition.
Therefore, it is of interest to study other driven systems with friction and explore the singularities of the
corresponding nonequilibrium steady states.

In the present paper we consider a generalization of the Ising model with a discrete symmetry, 
the $q$-state Potts model~\cite{Wu},
which is equivalent to the Ising model for $q=2$ but displays quite different critical behaviors when $q$ is varied. In the equilibrium case there are detailed analytical and numerical
informations about the physical properties of this system. The model has been solved in 1D for any $q$, in the large-$q$ limit in 2D, as well as on a fully connected lattice. In 2D, exact results are available at the phase-transition point~\cite{baxter}, which is of second (first) order for $q \le q_c(2)=4$ ($q > 4$). In three dimensions (3D), the limiting value $q_c(3)<3$. 
Here we investigate the $q$-state Potts model with friction and study the behavior of the steady state
as well as the properties of the nonequilibrium phase transition for different values of $q$ as a function of the
velocity $v$ and the friction interaction $K_f$. In 1D, a nonequilibrium phase transition is present only
at $v=\infty$ where the
problem can be solved exactly. In 2D we study the quasi-static limit, $v \to 0$, in which case exact results
are known about the friction force and its singularity at the equilibrium phase transition point.
At the nonequilibrium case a numerically exact treatment is given for $v=\infty$ in the large-$q$ limit.
These results are compared with numerical simulations which are performed at finite $v$ and for different values
of $q$.

The structure of the paper is the following. The model is introduced in Sec.~\ref{sec:2}, and the dissipation and the
friction force in the low $v$ limit are studied in Sec.~\ref{sec:2.uj}. We treat the model in the
limit $v=\infty$ in Sec.~\ref{sec:3}. Numerical simulations performed in the 2D case for $v$ and $q$ finite are presented in Sec.~\ref{sec:4}
and the results are discussed in Sec.~\ref{sec:5}. The solution of the self-consistency equation for the magnetization in 1D is given in Appendix~A and details about the calculation of the equilibrium surface magnetization
of the 2D Potts model in the large-$q$ limit are given in Appendix~B.

\section{Model}
\label{sec:2}

We consider two identical Potts models defined by the Hamiltonians ${\cal H}^{(s)}$ and ${\cal H}^{(s')}$,
respectively, which are coupled at their free surface by a time-dependent interaction term ${\cal V}(t)$.
The reduced Hamiltonian of the composite system takes the form
\be
\beta{\cal H}(t)=\beta{\cal H}^{(s)}+\beta{\cal H}^{(s')}+\beta{\cal V}(t)\,,
\ee
with $\beta=1/k_B T$.


\begin{figure}
\vglue 5mm
\begin{center}
\includegraphics[width=8.cm,angle=0]{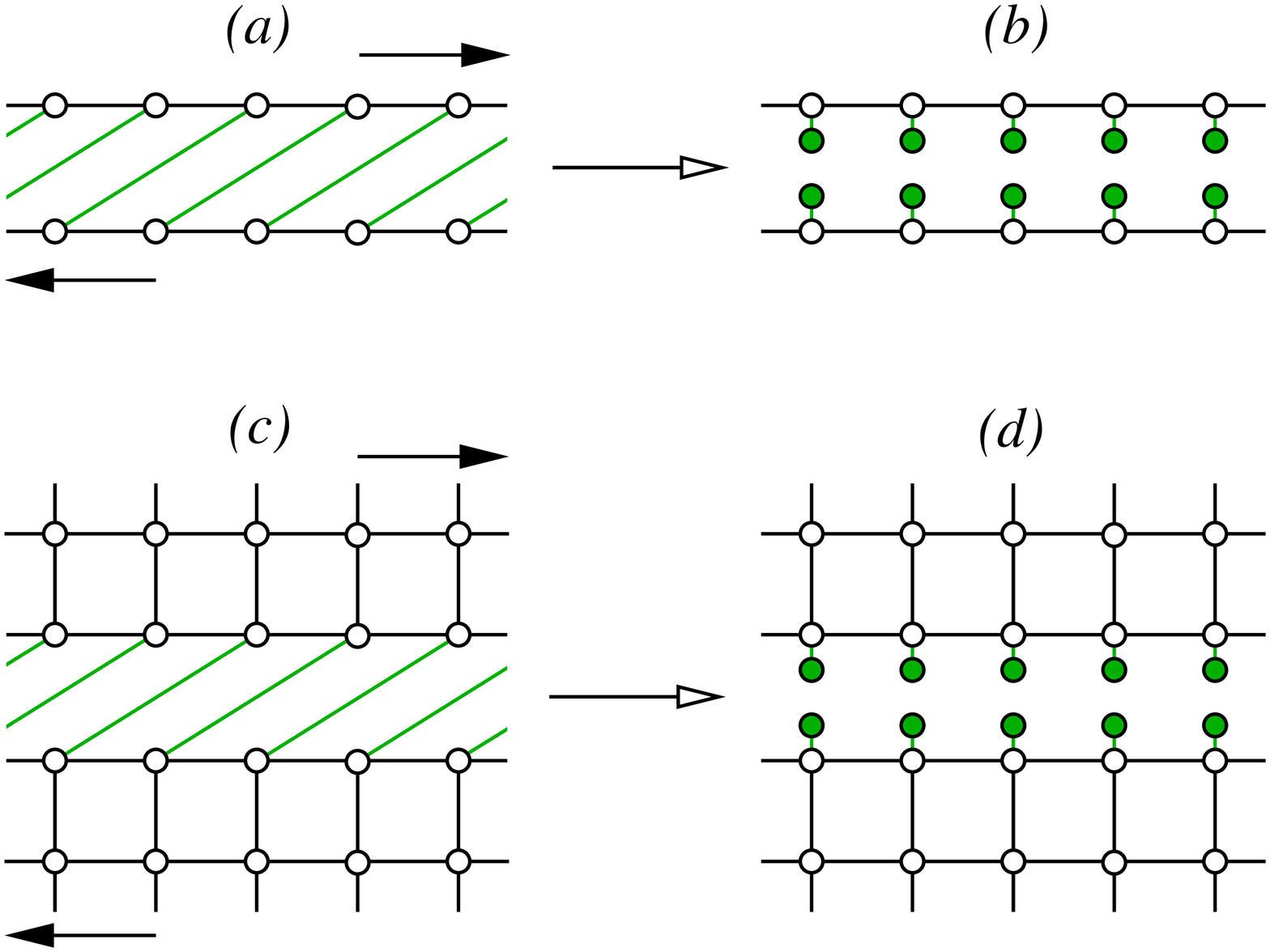}
\end{center}
\caption{(Color online) Driven interacting Potts models sliding on each other in (a) 1D and (c) 2D, with a translation by two lattice constants between interacting sites. For infinite relative velocities, the
system can be replaced by two decoupled Potts models in (b) 1D and (d) 2D, interacting with a set of fluctuating variables, attached to one layer of ghost sites [green (gray) circles]. These fluctuating variables  are equivalent to an effective surface field with strength $h_f$.} 
\label{fig1}
\end{figure}


In 1D [Fig.~1(a)] we have
\be
\beta{\cal H}^{(s)}=-K \sum_{i=1}^{N} \delta(s_i-s_{i+1})\,,
\label{H1D}
\ee
in terms of the Potts spin variables $s_i=0,1,\dots,q-1$ and $\delta(n)$ is the Kronecker delta function. In 
$\beta{\cal H}^{(s')}$, $s_i$ is replaced by $s'_i$. The interaction term is given by
\be
\beta{\cal V}(t)=-K_f\sum_{i=1}^N \delta(s_i-s'_{i+\Delta(t)})\,,
\label{V1D}
\ee
where $\Delta(t)=vt$, and periodic boundary conditions are used, $i+N\equiv i$.

In 2D [Fig.~1(c)] the Potts variable $s(i,j)$ are attached to the sites $(i,j)$ of a square lattice.
The Hamiltonian of the Potts model is then
\bea
\beta{\cal H}^{(s)}&=&-K_1 \sum_{i=1}^{N} \sum_{j=1}^L\delta(s_{i+1,j}-s_{i,j})\nonumber\\ 
&-&K_2\sum_{i=1}^{N} \sum_{j=1}^{L-1}\delta(s_{i,j+1}-s_{i,j})\,,
\label{H2D}
\eea
and similarly for $\beta{\cal H}^{(s')}$. The couplings in the horizontal and in the vertical directions, $K_1$ and $K_2$, can be different and the interaction term takes the form:
\be
\beta{\cal V}(t)=-K_f \sum_{i=1}^{N}\delta(s_{i,1}-s'_{i+\Delta(t),1})\,.
\label{V2D}
\ee
We set periodic boundary conditions in the horizontal direction, $s_{i+N,j} \equiv s_{i,j}$ as well as in the
vertical direction, which means that there are two equivalent sliding interfaces in the system.
The problem is generally studied in the thermodynamic limit. Note that the
2D problem with $L=1$ (i.e., with one layer in each subsystem) is formally equivalent to the 1D problem with $s(i)\equiv s(i,1)$.

\section{Dissipation and the friction force for $v \to 0$}
\label{sec:2.uj}
\begin{figure}
\vglue 5mm
\begin{center}
\includegraphics[width=8.cm,angle=0]{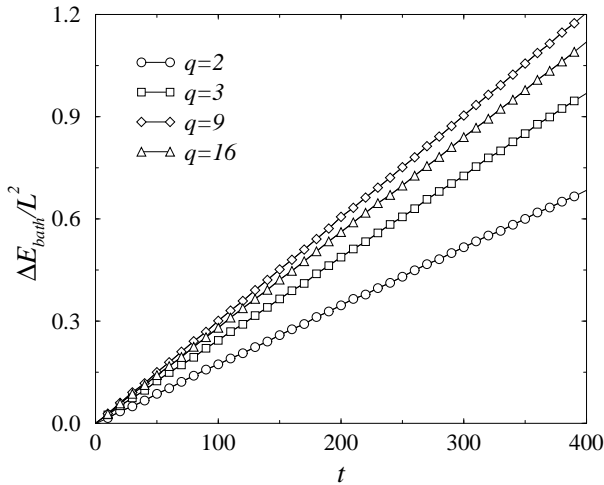}
\end{center}
\caption{Accumulated energy per spin, which is dissipated by the heat bath during time $t$
in a $80 \times 80$ lattice for different values of $q$. The sliding velocity is $v=1$ and the temperature is $T=1.1T_c(q)$,
see the text.} 
\label{fig7a}
\end{figure}

To see the relation of the above model with magnetic friction we follow Ref.~\cite{kadau},
couple the system to a heat bath of constant temperature, $T$, and study its nonequilibrium properties
by Monte Carlo simulations. In our case
the relaxation kinetics is governed by the heat-bath algorithm \cite{landau_binder}. Measuring the energy difference,
$\Delta E=E'-E$, between the  original ($E$) and the flipped ($E'$) configurations it is
found that $ \Delta E<0$ ( i.e. energy is dissipated to the heat bath, which shows the presence of magnetic
friction in our system). More detailed investigations are performed for the 2D problem with $L \times L$
spins (i.e., for $N=L$ with $L \times L/2$ subsystems) and with periodic boundary conditions, in which case the dissipated energy per spin during time $t$, $\Delta E_\mathrm{bath}(t)/L^2$, has been measured for different relative velocities, $v$, and temperatures, $T$. We illustrate
the time dependence of $\Delta E_\mathrm{bath}(t)/L^2$ in Fig.~\ref{fig7a} for $v=1$ and for $K_f=K_1=K_2=1/T$ at a temperature
$T=1.1T_c(q)$, above the bulk phase transition point, $T_c(q)=1/[\ln(1+\sqrt{q})]$, for different values of $q$. As for the Ising model \cite{kadau} with $q=2$ the dissipated energy grows linearly in time,
$\Delta E_\mathrm{bath}(t)=Pt$, and the dissipation rate, $P$, depends linearly on the velocity for small $v$:
$P=Fv$, where $F$ is the friction force. $F$ is found proportional to the length of the cut, $L$, so
that the magnetic frictional shear stress $F/L$ can be used to characterize the magnetic friction. 


\begin{figure}
\vglue 5mm
\begin{center}
\includegraphics[width=8.cm,angle=0]{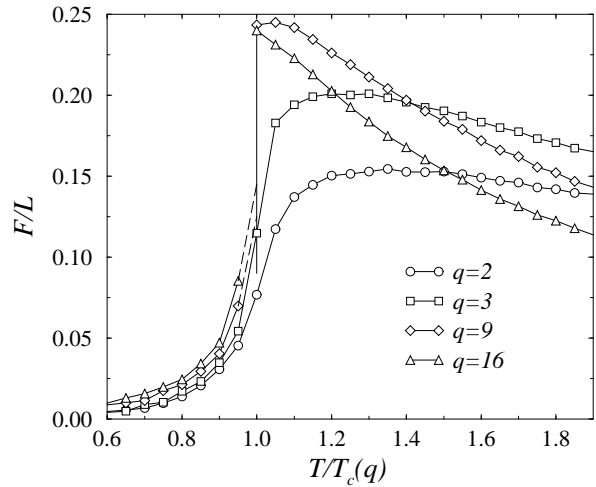}
\end{center}
\caption{The magnetic frictional shear stress, $F/L$, as a function of temperature
in a $80 \times 80$ lattice for different values of $q$ and a sliding velocity $v=1$. In the thermodynamic limit for $q=2$ and $3$
the derivative of $F/L$ is singular at $T_c(q)$, whereas for $q=9$ and $16$, $F/L$ has a discontinuity, which
is also indicated in the figure.} 
\label{fig8}
\end{figure}

We have measured the magnetic frictional shear stress at different temperatures. Data are presented
in Fig.~\ref{fig8} for $q=2,3,9$ and $16$. In a finite system with $80 \times 80$ spins $F/L$ shows some kind of
extremal behavior in the vicinity of the bulk phase-transition temperature, $T=T_c(q)$. For $q=2$ and $3$,
having a second-order equilibrium transition,
the derivative of $F/L$ with respect to $T$ shows a maximum, whereas for $q=9$ and $q=16$, having a first-order
equilibrium transition, $F/L$ itself is maximal around that point. In reality one is interested in the behavior
of the system in the thermodynamical limit, in which case we define $f=\lim_{L \to \infty}F/L$. 
In the limit $v \to 0$, the sliding velocity is so slow that the system has time to relax to equilibrium 
between successive relative moves of the two subsystems. Then
one can express $f$ as the difference between two equilibrium spin-spin correlation functions \cite{kadau}.
In the original configuration $E/L$ is proportional to the nearest-neighbor correlation function, $C_1=\langle\delta(s_{i,1}-s'_{i,1})\rangle$, and after the displacement $E'/L$ is proportional to the
next-nearest-neighbor correlation function, $C_2=\langle\delta(s_{i,1}-s'_{i,2})\rangle$,
so that the magnetic frictional shear stress is given by $f=(C_2-C_1)/2$, where the division by 2 is due to
the two equivalent sliding surfaces. For the Ising
model these correlations are known \cite{mccoy_wu} and thus $f(T)$ can be calculated exactly. Its derivative at $T=T_c$ shows a
logarithmic singularity:
\be
\left.\dfrac{{\rm d} f}{{\rm d} T}\right|_{T_c}\simeq \dfrac{2K_c}{\pi T_c}(1-\sqrt{2}) \ln\left|1-\dfrac{T}{T_c}\right|\;.
\label{f_sing_2}
\ee
Comparing this result with the numerical findings in Fig.~\ref{fig8} we can say that in the thermodynamic limit the slope of the curve at $T=T_c$ and $q=2$
is divergent for $v \to 0$. We expect that some kind of divergency will stay for a finite $v$ also
and the finite slope in Fig.~\ref{fig8} is a finite-size effect. One can
generalize this result for $q=3$ and more generally for systems having a second-order equilibrium transition.
Since the near-neighbor correlations have the same type of singularity as the energy density, we obtain
in the $v \to 0$ limit:
\be
\left.\dfrac{{\rm d} f}{{\rm d} T}\right|_{T_c}\sim \left|1-\dfrac{T}{T_c}\right|^{-\alpha}\;,
\label{f_sing_q}
\ee
where $\alpha$ is the specific heat critical exponent of the system (for the $q=3$ Potts model it is
$\alpha=1/3$). On the contrary, for $q=9$ and $q=16$ and more generally for systems with a first-order equilibrium
transition, the energy density, and thus the magnetic frictional shear stress has a discontinuity at $T=T_c$.
Consequently the true behavior in Fig.~\ref{fig8} for infinite systems is a jump for $q=9$ and $16$, at least in the
slow displacement ($v \to 0$) limit. This limiting behavior is indicated by the dashed and vertical lines in Fig.~\ref{fig8}.


In the experimentally relevant situation the sliding velocity is finite and the system under investigation
is out of equilibrium. In this case, as already demonstrated for the Ising model \cite{hucht}, a nonequilibrium phase transition
takes place in the system. 
The nonanalytical behavior of
the friction force in Fig. 3 strongly suggests the existence of a
nonequilibrium phase transition for $q>2$ as well.
In the following we study this nonequilibrium phase transition, first in the infinite velocity limit, in which case the mean-field
treatment is exact, and afterwards for finite $v$ by Monte Carlo (MC) simulations.

\section{Solution at infinite relative velocities}
\label{sec:3}

\subsection{Potts model with a fluctuating variable}
\label{sec:3.1} 
The solution of the problem becomes simple at $v = \infty$ as noticed already for the Ising model in Ref.~\cite{hucht}.
Here we use a generalization of the same argument for the $q$-state Potts model. The basic observation is that
at $v = \infty$, the spin variables $s_{i,1}$ and $s'_{i+\Delta(t),1}$  in the interaction term of Eqs.~(\ref{V1D}) 
and~(\ref{V2D}) are uncorrelated. Consequently, for the subsystem $(s)$, the spin $s'_{i+\Delta(t)}$ can be replaced by a randomly
chosen spin from the surface layer of the subsystem $(s')$. The same effect on $s_{i,1}$ is obtained if the surface spin is coupled to a fluctuating Potts variable $\mu_{i}=0,1,\dots,q-1$, with the following constraints on the mean-values
\be
\langle \delta(\mu_{i}\!-\!\alpha) \rangle\!\!=\!\!\langle \delta(s'_{i,1}\!-\!\alpha) \rangle\!=\dfrac{[\delta(\alpha)q-1]m_f+1}{q},
\label{mu}
\ee
where $m_f$ is the mean value of the magnetization  in the two interface layers and $\alpha=0,1,\dots,q-1$ is a Potts variable. Thus the interaction
term in Eq.~(\ref{V2D}) can be replaced by the following effective interaction for the subsystem $(s)$ (see Fig.~\ref{fig1})
\be
\delta(s_{i,1}-s'_{i+\Delta(t),1}) \to \delta(s_{i,1}-\mu_{i})\,,
\ee
with the probability distribution 
\be
p(\mu_i)=\dfrac{1-m_f}{q}+\delta(\mu_i)m_f
\label{pmui}
\ee
for the fluctuating variable. 

In the next step we integrate out the fluctuating variable $\mu_{i}$ and replace
its effect by an external field acting on $s_{i,1}$.

\subsection{Effective external field}
\label{sec:3.2} 

Here we consider a Potts model on a general lattice, with Hamiltonian ${\cal H}_{0}$ involving the Potts variables $s_k$. These 
variables are interacting with a set of fluctuating  Potts  variables $\mu_k$, through the couplings $K_k$. The total reduced Hamiltonian is given by
\be
\beta{\cal H}_{\mu}=\beta {\cal H}_{0} - \sum_k K_k \delta(s_k-\mu_k)\,,
\ee
where the $\mu_k$ are distributed as in Eq.~(\ref{pmui}) with
%
%
a mean value, $m_k$.
%
%
%
%

The fluctuating variables $\mu_k$ can be traced out from the partition function
${\cal Z_\mu}$. Making use of the identity
\be
e^{K_k \delta(s_k-\mu_k)}=1+(e^{K_k}-1)\delta(s_k-\mu_k)\,,
\ee
one obtains
\bea
{\cal Z_\mu}&=&{\rm Tr}_{s,\mu} e^{-\beta{\cal H}_{\mu}}={\rm Tr}_{s}e^{-\beta{\cal H}_{0}}{\rm Tr}_{\mu}\prod_k
e^{K_k \delta(s_k-\mu_k)}\nonumber\\ 
&=&\prod_k c(K_k,m_k)\nonumber\\
&&\ \ \ \ \times {\rm Tr}_{s}e^{-\beta{\cal H}_{0}}\prod_k \left[ 1
+d(K_k,m_k)\delta(s_k)\right]\,,
\label{Zmu}
\eea
where:
\bea
c(K_k,m_k)&=&\left[ 1+(e^{K_k}-1)\dfrac{1-m_k}{q}\right]\nonumber\\
d(K_k,m_k)&=&\dfrac{q(e^{K_k}-1)m_k}{q+(e^{K_k}-1)(1-m_k)}\,.
\eea

The expression for the partition function in Eq.~(\ref{Zmu}) can be compared
to that of a Potts model, coupled to a  static magnetic field $h_k$,
and defined by the Hamiltonian:
\be
\beta{\cal H}_{\rm eq}=\beta{\cal H}_{0}-\sum_k h_k \delta(s_k)\,.
\ee
For this model the partition function reads:
\bea
{\cal Z}_{\rm eq}&=&{\rm Tr}_{s} e^{-\beta{\cal H}_{\rm eq}}={\rm Tr}_{s}e^{-\beta{\cal H}_{0}}\prod_k
e^{h_k \delta(s_k)}\nonumber\\ 
&=&{\rm Tr}_{s}e^{-\beta{\cal H}_{0}}\prod_k \left[ 1
+(e^{h_k}-1)\delta(s_k)\right]\,.
\label{Zeq}
\eea
If we fix the values of the field $h_k$ to $\tilde{h}_k$ such that $e^{\tilde{h}_k}-1=d(K_k,m_k)$, that is
\be
\tilde{h}_k=\ln\left\lbrace 1+\dfrac{q(e^{K_k}-1)m_k}{q+(e^{K_k}-1)(1-m_k)}\right\rbrace\,,
\label{hk}
\ee
then the partition function in Eq.~(\ref{Zmu}) can be expressed with the
equilibrium partition function as:
\be
{\cal Z_\mu}=\prod_k c(K_k,m_k) \times {\cal Z}_{\rm eq}(\tilde{h})
\ee

\subsection{Application to the driven Potts model}
\label{sec:3.3} 

Now we turn back to our original problem where two interacting Potts models are moving
with a constant relative velocity $v$ and have a magnetization $m_f$ at the interface in the stationary state. 
In the limit $v=\infty$ this system can be replaced by two noninteracting Potts models interacting with a set of fluctuating variables, attached to one layer of ghost sites as shown in Fig.~\ref{fig1}. Integrating out the degrees
of freedom associated with the ghost sites, the effect of one subsystem on the other is equivalent to a static effective surface field with strength $h_f$. Using the formalism of Sec .~\ref{sec:3.2}, we have
$m_k=m_f$ and $K_k=K_f$ for all surface sites $k$, whereas ${\cal H}_0$ is the 
Hamiltonian of the subsystem defined in Eqs.~(\ref{H1D}) and~(\ref{H2D}).
The effective surface field is also the same for all surface sites, $\tilde{h}_k=h_f$, and follows from Eq.~(\ref{hk}) so that
\bea
e^{h_f}&=&1+\Omega(m_f,K_f)\nonumber\\
\Omega(m_f,K_f)&=&\dfrac{q\tau(K_f)m_f}{1-\tau(K_f)m_f}\,,
\label{Omega}
\eea
with
\be
\tau(K_f)=\dfrac{e^{K_f}-1}{e^{K_f}+q-1}\,.
\label{tau}
\ee
The surface magnetization in the equilibrium system is obtained as
\be
m_{f,{\rm eq}}=\dfrac{\dfrac{q}{N}\dfrac{\partial\ln {\cal Z}_{\rm eq}}{\partial h_f}-1}{q-1}\,,
\label{m_f}
\ee
and must satisfy the self-consistency equation
\be
m_{f,{\rm eq}}\left[h_f(K_f,m_f) \right] =m_f\,,
\label{self-cons}
\ee
with the appropriate values of the subsystem interactions.

When the transition is of second order, the transition point satisfies the condition $\partial m_{f,{\rm eq}}/\partial m_f|_{m_f=0}=1$~\cite{hucht}.
Using Eqs.~(\ref{Omega}) and~(\ref{self-cons}) we arrive at
\be
\chi^{(0)}_{f,{\rm eq}}\big|_c \,q\,\tau(K_{fc})=1\,,
\label{crit_cond} 
\ee
where $\chi^{(0)}_{f,{\rm eq}}=\partial m_{f,{\rm eq}}/\partial h_f|_{h_f=0}$ is the zero-field surface
susceptibility of the $q$-state Potts model.

\subsection{Analytical solution in 1D}
\label{sec:1d}
The magnetization of the 1D Potts model in the presence of an external field $h_f$
can be calculated by the transfer-matrix method. This is explained in Appendix~\ref{app:1} where, using
the relation of Eq.~(\ref{Omega}) between the magnetization $m_f$ and the effective field $h_f$, a closed
form for the self-consistency condition in Eq.~(\ref{self-cons}) is derived. Besides the
trivial solution
\be
m_f^{(0)}=0\,,
\ee
the other solutions are given by two roots of the cubic equation

\be
a_3\,m_f^3+a_2\,m_f^2+a_1\,m_f+a_0=0\,,
\label{polynom}
\ee
with coefficients:
\bea
a_0&=&-(q-2)\left[2\tau\left(e^K-1\right)+q(\tau-1) \right]\nonumber\\
a_1&=&-2\tau (q-2)^2\left(e^K-1\right) -q(q-1)\nonumber\\
&+& \tau^2\left\lbrace \left[ q\left(e^K-1\right)+4 (q-1)\right] \left(e^K-1\right)+q(q-1) \right\rbrace \nonumber\\
a_2&=&\tau(q-2)\left(e^K-1\right) \nonumber\\ 
&\times & \left\lbrace \tau\left[q\left(e^K-1\right)+4(q-1)\right]+2(q-1)
\right\rbrace\nonumber\\
a_3&=&-(q-1)\tau^2\left(e^K-1\right)\left[ q\left(e^K-1\right)+4(q-1)\right].
\label{ai}
\eea
As explained in Appendix~\ref{app:1}, the third root does not satisfy
the self-consistency condition in Eq.~(\ref{selfc_eq}). We have discarded this nonphysical root after
a direct substitution into Eq.~(\ref{selfc_eq}).

\begin{figure}
\vglue 5mm
\begin{center}
\includegraphics[width=8.cm,angle=0]{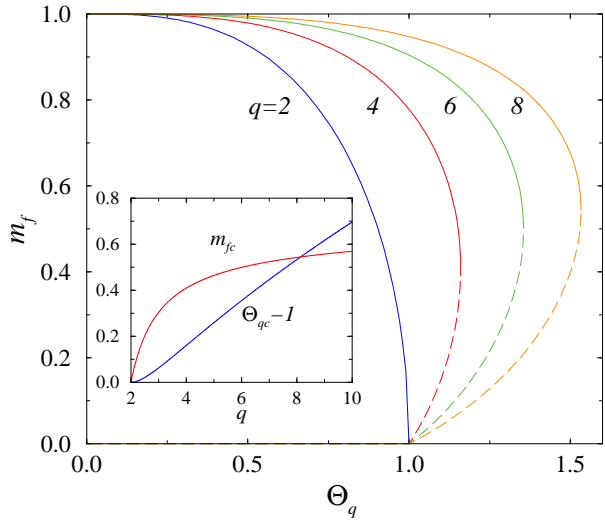}
\end{center}
\caption{(Color online) Temperature dependence of the magnetization $m_f$ of the $q$-state Potts chain with
friction in the limit $v=\infty$ for the symmetric model with $K=K_f$. The stable (unstable) solutions
of the self-consistency equation are denoted by full (broken) lines.
The transition is of second order for $q=2$ (Ising model) and first order for $q>2$. In the latter case
there is a hysteresis: on heating (cooling) the transition is at $\Theta_{qc}$ ($\Theta_{q1}=1$).
The inset gives the deviation of the transition point $\Theta_{qc}$ from the Ising value $\Theta_{2c}=1$,
as well as the value of the jump in the magnetization on heating, $m_{fc}$, as a function of $q$.}
\label{fig2}
\end{figure}

The structure of the cubic polynomial is different for $q=2$ (Ising model) and for $q>2$.
For the Ising model, due to the up-down symmetry, the even coefficients in Eq.~(\ref{polynom}) are
vanishing, $a_0=a_2=0$, and the nontrivial solutions are:
\be
m_f^{(\pm)}=\pm \sqrt{\dfrac{\tau^2e^{2K}-1}{\tau^2(e^{2K}-1)}},\qquad K \ge K_c\,.
\ee
We have checked that these solutions are the stable ones below the 
critical temperature $T_c$ which is given by the condition:
\be
\tau(K_{fc}) e^{K_c}=1\,.
\label{Ising_crpt}
\ee
Note that for the Ising model $\tau(K_f)=\tanh(K_f/2)$, thus we recover the result previously obtained in
Ref.~\cite{hucht}. As the critical point is approached
the nonequilibrium magnetization is vanishing continuously with a critical exponent $\beta(q=2)=1/2$.
The temperature dependence of the nonequilibrium magnetization for the symmetric case, $K_f=K$, is shown in Fig.~\ref{fig2}.
Here we use the temperature parameter $\Theta_q$, defined as:
\be
\Theta_q=\dfrac{q}{\sqrt{2}\left( e^{K}-1\right)}\,.
\label{thetaq}
\ee
For the symmetric Ising model at the critical point we have $\Theta_{2c}=1$ [see Eq.~(\ref{Ising_crpt})].

Now we turn to the solution of the non-Ising case, $q>2$, looking for
those roots of the cubic polynomial in Eq.~(\ref{polynom}) which are the nontrivial solutions
of the self-consistency equation. We know from Cardano's formula that the structure of the real solutions
of a cubic polynomial depends on a discriminant defined as ${\rm Disc}=Q^2+P^3$, with:
\bea
P&=&-\dfrac{1}{9}\left( \dfrac{a_2}{a_3}\right)^2+\dfrac{a_1}{3a_3}\nonumber\\ 
Q&=&\dfrac{1}{27}\left( \dfrac{a_2}{a_3}\right)^3-\dfrac{a_1a_2}{6a_3^2}+\dfrac{a_0}{2a_3}\,.
\eea
For ${\rm Disc}>0$, which happens when $\Theta_q>\Theta_{qc}$, the polynomial has one real root and there is no nontrivial
solution of the self-consistency equation. On the contrary, ${\rm Disc}<0$ in the low temperature region such that $0<\Theta_q<\Theta_{qc}$ and the polynomial has three real roots. For the two nontrivial solutions, such that $m_f^{(+)}>m_f^{(-)}$, we have checked that
$m_f^{(+)}>0$ in the whole region whereas $m_f^{(-)}<0$ ($m_f^{(-)}>0$) for $\Theta_q<\Theta_{q1}$
($\Theta_{q1}<\Theta_q<\Theta_{qc}$). Thus $\Theta_{q1}$ is defined by the condition $m_f^{(-)}=0$.
The two nontrivial solutions annihilate at the transition point $\Theta_{qc}$.

We have studied the stability of the solutions as well
as their domains of attraction by considering the self-consistency Eq.~(\ref{selfc_eq}) and varying $m_f$ on
its right-hand side. The solution $m_f^{(-)}$ is always unstable, on the contrary $m_f^{(+)}$  is always stable. The
region of attraction of the latter solution is $1>m_f>0$ ($1>m_f>m_f^{(-)}$) for $\Theta_q<\Theta_{q1}$
($\Theta_{q1}<\Theta_q<\Theta_{qc}$).
The trivial root $m_f^{(0)}=0$ is unstable for $\Theta_q<\Theta_{q1}$
and stable for $\Theta_q>\Theta_{q1}$  with the region of attraction
$m_f<m_f^{(-)}$ ($0\le m_f \le 1$) for $\Theta_{q1}<\Theta_q \le \Theta_{qc}$ 
($\Theta_q > \Theta_{qc}$). 


\begin{figure}
\vglue 5mm
\begin{center}
\includegraphics[width=8.cm,angle=0]{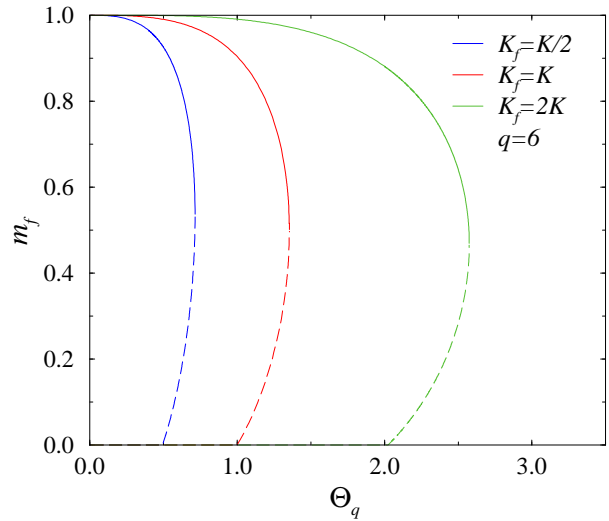}
\end{center}
\caption{(Color online) Influence of the coupling $K_f$ between the two chains on the temperature dependence of $m_f$ for $q=6$. The stability of the ordered phase increases with $K_f$ (from left to right).}
\label{fig3}
\end{figure}


The magnetization of the symmetric model for different values of $q$ is shown in
Fig.~\ref{fig2} as a function of $\Theta_q$. 
For $q>2$ the nonequilibrium phase transition is first order and the magnetization shows a hysteresis. Starting
from the ordered phase with $\Theta_q<\Theta_{q1}$ and heating the
system, the magnetization remains given by the nontrivial stable solution $m_f^{(+)}>0$ until $\Theta_{qc}$ where
it jumps to $m_f^{(0)}=0$, the trivial solution. In the reverse process, cooling the system
from the disordered phase with $\Theta_q>\Theta_{qc}$, the magnetization remains vanishing ($m_f^{(0)}=0$) until
$\Theta_{q1}$ where it jumps to the nontrivial solution $m_f^{(+)}>0$.
The location of the transition point $\Theta_{qc}$ and the value of the magnetization jump $m_{fc}=m_f(\Theta_{qc}\big|_-)-m_f(\Theta_{qc}\big|_+)$ are shown in
the inset of Fig.~\ref{fig2} for the symmetric model as a function of $q>2$. 

For $q>2$, $\Theta_{q1}$ corresponds to a vanishing nontrivial solution $m_f^{(-)}$, thus to $a_0=0$ according to Eq.~(\ref{polynom}). Using Eqs.~(\ref{tau}) and~(\ref{thetaq}) to express $a_0$ in Eq.~(\ref{ai}) leads to:
\be
\left(1+\dfrac{q}{\sqrt{2}\Theta_{q1}}\right)^{K_f/K}=\left(1+\dfrac{q\Theta_{q1}}{\sqrt{2}}\right)\,.
\ee
It follows that in the symmetric case, $K_f/K=1$, one obtains $\Theta_{q1}=1$ as shown in Fig.~\ref{fig2}. When $K_f/K$ increases, the ordered phase becomes more stable and $\Theta_{q1}$ increases, too.

The influence of the coupling ratio $K_f/K$ on the temperature dependence of the nonequilibrium magnetization is shown in Fig.~\ref{fig3} for $q=6$.

We have the following asymptotics for the symmetric model.
When $q$ is close to 2, the critical point and the magnetization discontinuity behave as: 
\be
\Theta_{qc} \approx 1+\dfrac{1}{8}(q-2)^2\,,\qquad  m_{fc}\sim q-2\,.
\ee
When $\Theta_q<\Theta_{qc}$, the magnetization approaches the limiting value $m_{fc}$ with a square-root singularity:
\be
m_f-m_{fc} \sim \sqrt{\Theta_{qc}-\Theta_q}\,.
\ee

For large-$q$ values the transition point is located at
\be
\Theta_{qc}\approx\sqrt{\dfrac{q}{2}}\left[1-aq^{-1/6} \right],\quad e^{K_c}\approx\sqrt{q}\left[1+aq^{-1/6} \right]\,,
\ee
and the magnetization discontinuity is given by
\be
m_{fc}\approx 1-bq^{-1/6}\,,
\ee
with $a\approx0.9447$ and $b\approx0.629$, thus $a/b\approx 3/2$.

\subsection{Solution in 2D for large-$q$ values}
\label{sec:2d}
The key point of the solution of the driven system in the limit $v=\infty$ is the knowledge of the
equilibrium surface magnetization as a function of temperature and surface field. For the Potts model in 2D, analytic results about the surface magnetization are known for $q=2$~\cite{mccoy_wu} (Ising model) as well as in the large-$q$ limit~\cite{ic99,cissz99,ti02}.
The analysis for the Ising model has been performed in Ref.~\cite{hucht} and here we consider
the Potts model in the large-$q$ limit.

\begin{figure}
\vglue 5mm
\begin{center}
\includegraphics[width=8.cm,angle=0]{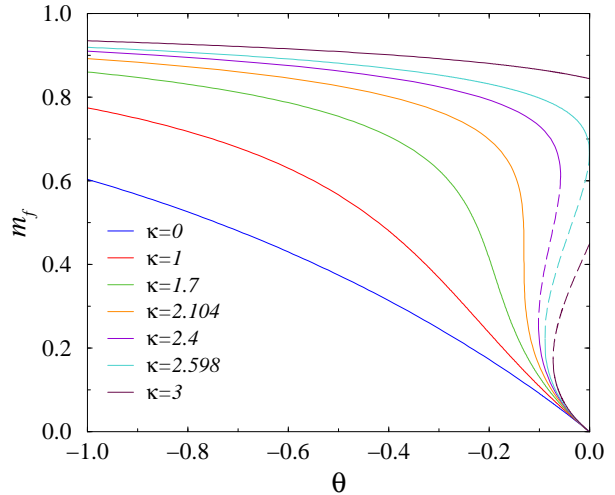}
\end{center}
\caption{(Color online) Temperature dependence of the interface magnetization of the 2D Potts model with
friction in the $v=\infty$ and large-$q$ limits, using the strongly anisotropic (Hamiltonian) version
of the problem, for values of the interface coupling $\kappa$ increasing from left to right. When the strength of the interface coupling increases, there is a critical value $\kappa_c'$ above which the system becomes bistable with an unstable solution (broken line) between two stable ones (full lines). The interface magnetization always vanishes in the bulk disordered phase, $\theta>0$.}
\label{fig4}
\end{figure}

The 2D Potts model is defined in Eq.~(\ref{H2D}) with the time-dependent interaction term given in Eq.~(\ref{V2D}).
We treat the problem in the strongly anisotropic limit~\cite{kogut} where the horizontal coupling $K_1\to\infty$, the vertical coupling $K_2\to0$, while the ratio $J=K_2/K_1^*$ remains constant. Here $K_1^*$ is the dual coupling defined through:
\be
\left(e^{K_1}-1\right) \left(e^{K_1^*}-1\right)=q\,.
\ee
Then the column-to-column transfer matrix of the noninteracting system in Eq.~(\ref{H2D})
takes the form ${\bf T}=\exp(-K_1^* {\bf H})$, where ${\bf H}$ is the quantum Hamiltonian~\cite{solyom}:
\be
{\bf H}=-J\sum_{j=1}^{L-1} \delta(s_j-s_{j+1})-\dfrac{1}{q}\sum_{j=1}^{L} \sum_{k=1}^{q-1} {\bf M}_j^k\,.
\label{H_op} 
\ee
In the last term, ${\bf M}_j$ is a spin-flip operator such that ${\bf M}_j^k s_j=s_j+k,{\rm mod}(q)$.
The quantum Potts model defined in Eq.~(\ref{H_op}) has a quantum phase transition at $J=J_c=1$ in the thermodynamics limit.
This transition is of second order for $q \le 4$ and first order for $q>4$.
The interaction between the two driven systems in the limit $v=\infty$ is represented by a
static surface field, the strength of which is obtained from Eq.~(\ref{Omega}) as $h_f=K_f m_f$ in the strongly anisotropic limit, $K_f\to0$. Thus, using the parametrization $K_f=K_1^*\kappa/\sqrt{q}$, the Hamilton operator in Eq.~(\ref{H_op}) is supplemented
by a surface-field term:
\be
{\bf V}=-h\,\delta(s_1)\,,\qquad h=\dfrac{h_f}{K_1^*}=m_f\dfrac{\kappa}{\sqrt{q}}\,.
\label{V_s} 
\ee
The surface critical behavior of the quantum Potts model in the large-$q$ limit has been studied in Refs.~\cite{ic99,cissz99,ti02}
and the results are summarized in Appendix~\ref{app:2}. In the large-$q$ limit, corrections of the order of $q^{-1/2}$
are taken into account and the distance from the critical point is defined as $J_c-J=\theta/\sqrt{q}$,
where $\theta$ plays the role of a reduced temperature. Using these results we
can calculate the equilibrium surface magnetization $m_{f,{\rm eq}}[\theta,h_f(\kappa,m_f)]$ and the nonequilibrium magnetization is deduced from the self-consistency condition in Eq.~(\ref{self-cons}).
We have solved the self-consistency equation for the interface magnetization numerically exactly~\cite{numerics}. The dependence of $m_f$ on the reduced temperature $\theta$ for different values of the interface coupling $\kappa$ is shown in Fig.~\ref{fig4}.


For $\kappa=0$ we recover the magnetization at a {\it free surface}, which vanishes linearly at the bulk critical
temperature $\theta_c=0$. With increasing interaction $\kappa$ the interface magnetization increases for $\theta<0$,
but it is always vanishing in the bulk disordered phase, $\theta>0$. This behavior is due to the fact that the equilibrium surface
magnetization of the model $m_1=0$ above $\theta_c$, whatever the value of the surface field, as seen in Appendix~\ref{app:2}. This is a peculiarity of the system in the large-$q$ limit which will be discussed further in Sec.~\ref{sec:4.3}.

In the ordered phase, the shape of the magnetization curve qualitatively changes with increasing $\kappa$.
At a critical value $\kappa_c'=2.104$, its slope diverges when $\theta=\theta_c'=-0.13037$.
In the vicinity of this critical temperature the nonequilibrium interface magnetization shows a power-law singularity
$m(\kappa_c',\theta)-m(\kappa_c',\theta_c') \sim |\theta-\theta_c'|^{1/3}$. Increasing the strength of the
interface coupling further, $\kappa>\kappa_c'$, a bistability occurs. Here in a finite range of
temperature, $\theta_1 < \theta < \theta_2\leq0$,
there are three solutions of the self-consistency equation, which are denoted by $m_f^{(1)} < m_f^{(2)} < m_f^{(3)}$.
Among these $m_f^{(2)}$ is unstable, whereas $m_f^{(1)}$
is stable for perturbations in the region $m_f<m_f^{(2)}$ and $m_f^{(3)}$ is stable for $m_f>m_f^{(2)}$.
Consequently starting at sufficiently low temperature,
$\theta<\theta_1$ and heating the system, its interface magnetization will stay on the upper part of the curve and
follows the solution $m_f^{(3)}$ in the range $\theta_1<\theta<\theta_2$. At $\theta=\theta_2$ it jumps to the lower
part of the curve, which is the continuation of the solution $m_f^{(1)}$. This jump being finite the transition is of the first order. When $\theta_2<0$, heating the system further the interface magnetization vanishes at $\theta_c=0$ linearly. In the reverse process we start in the disordered phase, $\theta>0$, and cool down the system. Then the interface magnetization increases linearly below $\theta_c=0$ and follows  the solution $m_f^{(1)}$,
in the range $\theta_2>\theta>\theta_1$. At $\theta=\theta_1$ it jumps to
the upper part of the curve, which is the continuation of the solution $m_f^{(3)}$. Consequently there
is a hysteresis in the temperature dependence of the nonequilibrium magnetization.

The limiting value $\theta_2$ first increases with $\kappa$ until $\kappa=\kappa^*=\sqrt{\dfrac{27}{4}}\approx2.598$ above which it stays at $\theta_2=0$. For $\kappa>\kappa^*$ on heating the nonequilibrium interface
magnetization jumps directly from $m_f^{(3)}(0)>0$ to $m_f=0$ at $\theta=0$~\cite{theta=0}. Thus, on heating, for strong
interaction $\kappa >\kappa^*$, there is a first-order nonequilibrium transition at $\theta=0$.
In the reverse process, on cooling, the second-order transition at $\theta=0$ is followed by a first-order
one at $\theta=\theta_1$.

\section{Monte Carlo simulations of the 2D system}
\label{sec:4}

In the present section the results obtained previously in the limit $v=\infty$ are confronted with Monte Carlo simulations. In particular, we want to see how the fluctuations,
introduced by a finite relative velocity between the two driven systems, influence the properties
of the nonequilibrium interface magnetization and its singular behavior at the phase-transition point.
For the Ising model with $q=2$ these questions have been studied in Ref.~\cite{hucht}. In 1D the
ordered phase is suppressed for any finite value of $v$.
On the contrary in 2D the nonequilibrium fluctuations introduced by the finite velocity are found to be irrelevant and 
the nonequilibrium phase transition is described by the
same (mean-field) critical exponents.

Concerning the Potts model in 1D the nonequilibrium fluctuations should destroy
the ordered phase for any finite value of $q$, thus one does not expect any nonequilibrium phase transition for $v < \infty$.
However, the question is more delicate in 2D, where the properties of the nonequilibrium phase transitions
in the limit $v=\infty$ are different for $q=2$ and for large-$q$ values. These changes are likely to be
related to the surface-field dependence of the surface magnetization in the equilibrium systems
[see the self-consistency condition in Eq.~(\ref{self-cons})], which is known
to depend on the order of the bulk equilibrium phase transition. Therefore it is advisable to study the two regimes in the Monte Carlo simulations. Thus we treat successively the case $q=3$, where the equilibrium transition is second order, in Sec.~\ref{sec:4.2},
and the case $q=9$, where it is strongly first order, in Sec.~\ref{sec:4.3}.

\subsection{Method of simulation}
\label{sec:4.1}

We simulate systems consisting of $L \times L$ spins (i.e., for $N=L$ with $L \times L/2$ subsystems), with $L$ ranging from 80 to 320.
We use symmetric couplings between pairs of spins, with $K_1 = K_2 = 1/T$, but allow for a varying 
coupling strength $K_f$ across the cut separating our two Potts systems,
with $K_f$ ranging from $0.25/T$ to $2/T$. In the following, we denote the ratio $K_f/K_i=K_fT$ as $\kappa$.
In the horizontal direction (i.e., the direction parallel to the interface) periodic boundary
conditions are used. In the vertical direction we use both periodic boundary conditions 
(yielding two interfaces) and open boundary conditions and find that, in general, the local quantities
close to the interface are independent of the vertical boundary condition. Small deviations, if any,
only show up for the smallest system size. The data discussed in the following have been obtained
with periodic boundary conditions in both directions.

Our main focus is on the magnetization profile and, especially, on the magnetization close to the 
interface separating the two $q$-state Potts systems that move with the relative constant velocity $v$.
The magnetization of row $i$ is given by
\begin{equation}
\left< m(i) \right> = \left( q N_m/L -1 \right) / (q -1)
\end{equation}
where $N_m = \mathrm{max}(N_0, N_1, ..., N_q)$. Here $N_q$ is the average number of spins in state $q$ in row $i$.
Obviously, we have that $m_f = \left< m(1) \right> = \left< m(L/2)  \right>= \left< m(L/2+1)  \right>= \left< m(L) \right>$.

For the Monte Carlo updates we use the standard heat-bath algorithm as our single-spin flip algorithm. 
For the corresponding 2D Ising model~\cite{hucht} it was shown that qualitatively the results 
obtained in simulations are independent of the update scheme. This is different when looking at specific quantities,
as, for example, the phase transition temperature, which do depend on the chosen algorithm. In the following we
restrict ourselves to a qualitative discussion of the properties of our nonequilibrium system.

We implement the sliding of one half of the system with respect to the other half in the same way as in Refs.~\cite{kadau,hucht}. When simulating a system with sliding velocity $v$ we translate the upper half of our system
by one lattice constant after $L^2/v$ random sequential single spin updates. One Monte Carlo step therefore consists of
$L^2$ single spin flips and $v$ translations. In our numerical study, we varied $v$ between 1 and 80.

Based on the results discussed in the previous sections, we expect to observe a discontinuous change of the magnetization
close to the cut for $q > 4$. To observe a possible hysteresis we made both heating and cooling runs. Starting the
heating (cooling) runs with a fully ordered (disordered) initial state, we typically
let the system relax for 2 $10^5$ Monte Carlo steps at the first temperature before starting the measurement.
After averaging over typically $10^5$ steps, we changed the temperature and let the system relax for a few 10 000 
time steps before starting the measurement at the new temperature. We carefully monitored our system to detect
the possible presence of a hysteresis. This procedure was repeated at least ten times and the data discussed in this Section
result from averaging over these independent runs.

\subsection{Results for $q=3$}
\label{sec:4.2}


\begin{figure}
\vglue 5mm
\begin{center}
\includegraphics[width=8.cm,angle=0]{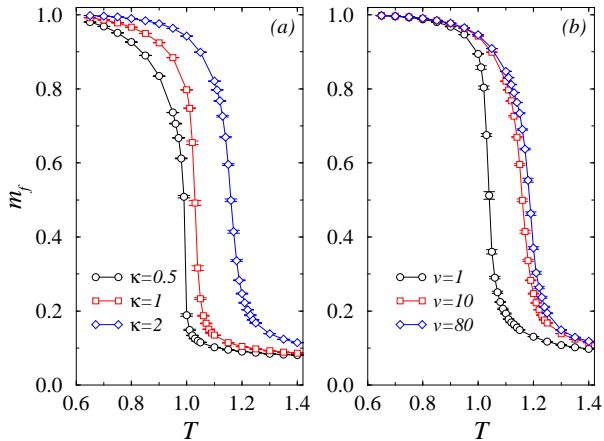}
\end{center}
\caption{(Color online) 
Interface magnetization for the $q=3$ model, with (a) $v= 10$ and various values of $\kappa$ and (b) $\kappa= 2$
and various values of $v$.
In all cases a continuous surface phase transition is observed.
The data shown here have been obtained for a system containing $160 \times 160$ spins. }
\label{fig5}
\end{figure}


The equilibrium $q =3$ Potts bulk system exhibits a continuous phase transition at the temperature
$T_c(q=3) = 1/\ln(1 + \sqrt{3})\approx 0.995$, similar to the Ising model which can be viewed as the $q=2$ Potts model.
We show in Fig.~\ref{fig5} the temperature dependence of the interface magnetization for various coupling strengths $\kappa$
and various sliding velocities $v$. In all cases we observe a continuous boundary phase transition at a critical 
temperature that depends on both $v$ and $\kappa$. Thus an increase of the coupling strength across the cut 
yields an increasing strength of the effective surface field that stabilizes the interface magnetization against 
thermal fluctuations. An increase of the boundary phase transition temperature is also observed when increasing
the sliding velocity, which is similar to what is observed for the Ising model~\cite{hucht}. In fact, the $q=3$
Potts model behaves in every aspect like the Ising model. Both models have a continuous bulk phase transition,
in both models the boundary phase transition is continuous and of mean-field type [we checked the mean-field character
of our transitions by studying the effective exponent $\beta_{\rm eff} = d\ln m_f/d\ln (T_c -T)$ and found that this
exponent tends to 1/2 when approaching $T_c$].

\subsection{Results for $q=9$}
\label{sec:4.3}


\begin{figure}
\vglue 5mm
\begin{center}
\includegraphics[width=8.cm,angle=0]{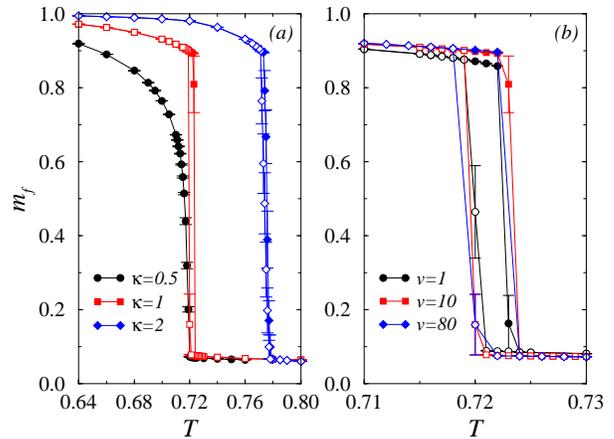}
\end{center}
\caption{(Color online) 
Interface magnetization for the $q=9$ model, with (a) $v= 10$ and various values of $\kappa$ and (b) $\kappa= 1$
and various values of $v$.
The order of the surface transition changes from continuous
for weak interface couplings to discontinuous for strong interface couplings. Note the changes of temperature scales.
The discontinuous character of the phase transition is revealed by the presence of a thermal hysteresis [filled (open) symbols on heating (cooling)].
}
\label{fig6}
\end{figure}


To see whether the scenario obtained for the $v=\infty$ and large-$q$ limits is generic for values of $q > 4$, we have
studied intensively the case $q =9$. For that value of $q$ the bulk system undergoes a strong first-order transition
at the temperature $T_c(q=9) = 1/\ln(1 + \sqrt{9}) \approx 0.721$. 

Our main findings are summarized in Fig.~\ref{fig6}. Fixing $v$ and changing the interface strength $\kappa$ reveals two
interesting features, see Fig.~\ref{fig6}(a). First we note that for small values of $\kappa$ the boundary phase transition
is continuous and takes place at a temperature that is comparable to the temperature of the bulk transition. For larger values
of $\kappa$, however, the boundary transition is discontinuous as revealed by a thermal hysteresis. We therefore
have the interesting situation that a continuous and a discontinuous surface transitions are separated by a tricritical point.
This behavior is in full agreement with the $v=\infty$ and large-$q$ scenario discussed in Sec.~IIID. There is, however, also
a remarkable difference between the $q = 9$ system and the large-$q$ case. Whereas in the large-$q$ case the interface 
magnetization is strictly zero above the equilibrium phase transition temperature, irrespective of the value of the interface coupling, for $q = 9$ we find for large values of $\kappa$ a finite surface magnetization above $T_c(q=9)$, where
the bulk is disordered, with a subsequent discontinuous surface transition at some temperature $T_s > T_c(q=9)$. This unexpected behavior
can be understood by mentioning that the corresponding equilibrium semi-infinite system has a finite surface magnetization
in strong enough fields for temperatures above $T_c(q=9)$~\cite{q_surf}.

Interestingly, the surface transition temperature for a fixed value of $\kappa$ is largely independent 
of the magnitude of the sliding velocity $v$. This is shown in Fig.~\ref{fig6}(b) for $\kappa = 1$. Obviously,
the limit $v=\infty$ is approached very rapidly and already modest values of $v$ yield results that are very
close to those expected for large values.

\section{Discussion}
\label{sec:5}
In this paper we have studied the magnetic contribution to the friction in the $q$-state Potts model
in which two interacting systems are moving with a constant relative velocity $v$. During the movement of the macroscopic
bodies there is a permanent energy flow into the heat bath, resulting from the friction force. The system is
thus driven into a nonequilibrium steady state, which can show order-disorder phase transitions as the temperature,
the strength of the interaction or the velocity are varied. The Potts model, with its rich critical behavior at equilibrium, depending on the value of $q$,
is a suitable system to study under nonequilibrium conditions.

We have studied the phase diagram and the phase transitions in this nonequilibrium system for different values of $q$
in 1D and 2D by analytical and numerical methods. In 1D long-range order is present only in the $v=\infty$
limit, where fluctuations are completely suppressed, and the problem is solved exactly using the mean-field method.
For finite $v$ in a system of finite extent, $N < \infty$, one expects cross-over phenomena in analogy to the
$q=2$ case \cite{hucht}. The nonequilibrium phase transition in this system is of second order for the
Ising model ($q=2$) but of first order
for $q>2$. In the latter case there is a hysteresis: the jump in the nonequilibrium magnetization takes place
at different temperatures on heating and cooling the system, respectively. This type of behavior is expected to take
place for other driven 1D models for $v=\infty$ also, provided the up-down symmetry of the local
order parameter is absent.

In 2D the nonequilibrium phase diagram is found to be more interesting and more exotic. Here the quasi-static
limit, $v \to 0$, is different from the true nonequilibrium case $v>0$. In the quasi-static limit the friction force has
the same type of singularity at the equilibrium phase transition point as the equilibrium energy density. On the
contrary, for any finite $v>0$ the singularity in the steady state at the phase transition point is controlled
by another fixed point, in which the critical exponents are mean-field like. More detailed results are obtained in the
large-$q$ limit, where for large-$v$ the problem is solved exactly using the mean-field method. 
Here the phase transition is of second order for sufficiently weak interaction. With increasing interaction a second phase transition takes place, at a lower temperature, which is continuous for a critical value of the coupling and discontinuous
for larger couplings. This latter transition is accompanied by a hysteresis. Numerical studies of the $q=9$
state Potts model with finite velocity have shown a similar scenario as described above for large~$q$.

On the contrary, numerical results obtained for the $q=3$ model have shown just one continuous transition for any value
of the couplings, which is of the mean-field type. This is similar to the scenario found for the Ising model. This difference in
the phase diagram is expected to be related to the nature of the corresponding equilibrium phase transition.
If the equilibrium phase transition is continuous, which happens for $q \leq 4$, the surface phase transition
is continuous also and the zero-field surface susceptibility is nonzero even above the bulk transition temperature.
In this case, a second-order nonequilibrium transition is expected to take place at a temperature which is
higher than the equilibrium bulk transition temperature. On the contrary, if the equilibrium phase transition
is first order, which is the case for $q>4$, the surface transition is usually second order, a phenomenon which
is known as {\it surface-induced disorder}~\cite{SID,SID0,SID1,SID2,SID3}. In this case the zero-field surface susceptibility
is zero at and
above the bulk transition temperature and one needs a finite surface field, $h_s > h_{sc}(q)>0$, to have
a nonvanishing surface magnetization at and above the transition point~\cite{q_surf}. 
This noncontinuous surface field dependence of the surface magnetization is 
responsible for the different scenario in the nonequilibrium system for 
$q>4$, in particular, for the first-order transition. The hysteresis, which 
accompanies the first-order transition, is due to the nonequilibrium nature of the process and the area of 
the hysteresis loop is proportional to the energy  dissipated during the transition.

Since the properties of surface-induced disorder are expected to have the same type of discontinuous
surface field dependence for any equilibrium first-order transitions in 2D and 3D~\cite{SID0}, the
same type of nonequilibrium scenario, which we have found for
the 2D Potts model with $q>4$, is expected to take place in these systems.

We close our paper with some remarks about the possible occurrence of nonequilibrium phases and phase transitions in other
driven systems. As we have seen in our study, there is an intimate connection between the equilibrium surface
critical behavior of these systems and the nonequilibrium states with friction. This relation is evident in the 
limit $v=\infty$, but fluctuations caused by a finite velocity are expected to be irrelevant, provided there is
a surface ordering in equilibrium. The surface critical behavior at equilibrium is a
complicated phenomenon~\cite{binder83,diehl86,pleimling04} in which one should take into account the effect
of enhanced or reduced surface
couplings and study the critical behavior at the different fixed points (ordinary, extraordinary, special, surface, etc.).
Also different considerations should be made for systems having an order parameter with continuous symmetry,
such as the $XY$- or the Heisenberg model~\cite{berche03}.
Finally, one could also think about using different geometries, such as edges, wedges~\cite{corner,igloi93}, or
parabolic shapes~\cite{peschel91,igloi93} to model a tip sliding on a flat surface.

\begin{acknowledgments}
This work has been supported by the Hungarian National Research Fund under grants No. OTKA K62588, K75324 and K77629 
(F. Igl\'{o}i), and by the US National Science Foundation Grant No. DMR-0904999 (M. Pleimling). The Institut Jean Lamour is Unit\'e Mixte de Recherche CNRS No. 7198.
\end{acknowledgments}

\appendix
\section{Self-consistency equation in 1D}
\label{app:1}
Here we consider the 1D Potts model in Eq.~(\ref{H1D}) in the presence of a field
\be
\beta{\cal H}^{(s)}=-K \sum_{i=1}^{N} \delta(s_i-s_{i+1})-h_f\sum_{i=1}^{N}\delta(s_i)
\label{H1D+}
\ee
and use the periodic boundary condition, $s_{N+1}\equiv s_1$. The transfer matrix of the problem is a $q \times q$
symmetric matrix ($q \ge 2$)
\be
{{\textbf T}} = \left(
\begin{matrix}
e^{K+h_f} & e^{h_f/2}   & e^{h_f/2}    &  \dots     &    e^{h_f/2}    \cr
  e^{h_f/2} &  e^K  &   1  &   \dots    &    1     \cr
  e^{h_f/2}  &  1  & e^K   &1 &       1  \cr
  \vdots  &  \vdots   & 1 &\ddots &      1  \cr
  e^{h_f/2} &  1   &  \dots    &   1   & e^K  \cr
\end{matrix}
\right)\,.
\label{T1D}
\ee
in terms of which the partition function reads ${\cal Z}_{\rm eq}={\rm Tr}\left\lbrace {\textbf T}^N\right\rbrace $. In
the large-$N$ limit, ${\cal Z}_{\rm eq}=\lambda_{m}^N$, where $\lambda_{m}$
is the leading eigenvalue of the transfer matrix given by:
\bea
\lambda_{m}&=&\dfrac{{e^{K+h_f}+e^{K}+q-2}}{2}\nonumber\\ 
&+&\dfrac{\sqrt{(e^{K+h_f}-e^{K}-q+2)^2+4 e^{h_f}(q-1) } }{2}\,.
\eea
Thus, we have
\be
\dfrac{\partial \ln \lambda_{m}}{\partial h_f}\!=\!
\dfrac{1}{2}\!\left[\!1\!+\!\dfrac{e^{K+h_f}-e^{K}-q+2}{\sqrt{(e^{K+h_f}\!\!-\!e^{K}\!\!-\!q\!+\!2)^2\!+\!4 e^{h_f}(q\!-\!1)}} \!\right]\,,
\label{lambda_der}
\ee
and the magnetization follows from Eq.~(\ref{m_f}).

Now taking the value of the static effective field in Eq.~(\ref{Omega}) and using~(\ref{lambda_der})
in Eq.~(\ref{m_f}) we obtain
\be
2(q\!-\!1)m_{f,{\rm eq}}\!-\!q\!+\!2\!=\!\dfrac{q(e^K \Omega-q+2)}{\sqrt{\left(e^K\Omega\!-\!q\!+\!2\right)^2\!+\!4(q\!-\!1)(\Omega\!+\!1)}}
\label{selfc_eq} 
\ee
where $\Omega=\Omega(m_f,K_f)$ is defined in Eq.~(\ref{Omega}). To obtain a self-consistent
solution, $m_{f,{\rm eq}}=m_f$, we first take the square of both sides of Eq.~(\ref{selfc_eq}) leading to
\bea
&&\!\!\!\!\!\!\!\left[(e^K\Omega\!-\!q\!+\!2)^2\!+\!4(q\!-\!1)(\Omega\!+\!1)\right]\nonumber\\ 
&&\times\left[2(q\!-\!1)m_f\!-\!q\!+\!2\right]^2\!=\!q^2(e^K\Omega\!-\!q\!+\!2)^2\,,
\label{square_eq}
\eea
which has always the trivial solution $m_f=0$. Then we
multiply Eq.~(\ref{square_eq}) by $(1-m_f\tau)^2/qm_f$ and obtain the cubic polynomial
which is given in Eq.~(\ref{polynom}).
Note that this polynomial has an extra root which is a solution of the squared equation, but not of the original one.

\section{Surface magnetization of the quantum Potts model for large-$q$ values}
\label{app:2}
In the following we consider the ground state of the quantum Potts model defined in Eq.~(\ref{H_op}), extended
by the surface field term in Eq.~(\ref{V_s}). We use fixed-spin boundary conditions at $j=L$ and the
surface at $j=1$ is free. In the large-$q$ limit, at the critical point $J_c=1$, the ground state of the
system is $L$-fold degenerate, having an energy $E_0^{(0)}=-J(L-1)$. In a first-order perturbative treatment~\cite{ic99,cissz99,ti02}, corrections
of the order of $1/\sqrt{q}$ are obtained through the solution of the following secular eigenvalue problem,
${\bf h}{\bf v}_\alpha=\epsilon_\alpha {\bf v}_\alpha$. Here ${\bf h}$ is a symmetric $L \times L$ matrix
\be
{\bf h} =-\dfrac{1}{\sqrt{q}} \left(
\begin{matrix}
h &  1  &      &       &     0   \cr
  1 &  \theta  &   1  &       &         \cr
    &  1  & 2\theta   &\ddots &         \cr
    &     &\ddots&\ddots &      1  \cr
  0 &     &      &   1   &(L-1)\theta   \cr
\end{matrix}
\right)\,,
\label{htilde}
\ee
where $\theta$ plays the role of a reduced temperature, since the coupling is parametrized as
$J = 1-\theta/\sqrt{q}$. Using the components of the ground-state eigenvector
$v_0(k)$ with $k = 1$, \ldots, $L$ one can express the 
magnetization profile as
\be
m_j=\sum_{k=1}^{j}\left[v_0(k)\right]^2\,,
\label{magnet1}
\ee
and the surface magnetization $m_1\equiv m_f$.

\begin{figure}
\vglue 5mm
\begin{center}
\includegraphics[width=8.cm,angle=0]{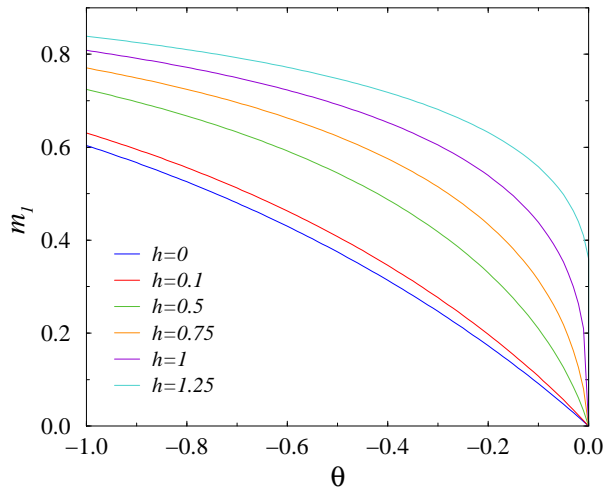}
\end{center}
\caption{(Color online) Surface magnetization of the large-$q$ state Potts model as a function of
the reduced temperature $\theta$ for different values of the surface field $h$. In the disordered phase,
$\theta>0$, the surface magnetization vanishes for any finite value of $h$. For the different curves $h$
increases from left to right.}
\label{fig7}
\end{figure}

The surface magnetization as a function of the reduced temperature $\theta$ is shown
in Fig.~\ref{fig7} for different values of the surface field $h$.

At the critical point, $\theta=0$, the surface magnetization is vanishing for $h \le 1$ and
it starts linearly for small $\theta$ as:
\be
m_1(\theta,h)=\dfrac{-\theta}{(1-h)^2}+O[\theta^2\xi(\theta)],\qquad h<1\,.
\ee
The second derivative, $\partial^2 m_1(\theta,h)/\partial \theta^2 \sim \xi(\theta)$, diverges as $\xi \sim \theta^{-1/3}$.
For $h > 1$, there is a finite surface magnetization at the critical point
and a small-$\theta$ behavior given by:
\be
m_1(\theta,h)=1-\dfrac{1}{h^2}+\dfrac{2h}{(h^2-1)^2}(-\theta)+O(\theta^2),\quad h>1\,.
\ee
In the disordered phase, $\theta>0$, the surface magnetization vanishes for any finite value of $h$.
Consequently, there is a first-order surface phase transition at $h=1$.


\begin{thebibliography}{99}
\bibitem{theta=0} At this temperature
the magnetizations  $m_f^{(2)}(0)$ and $m_f^{(3)}(0)$ are given by the positive roots of the equation
$m=1-(\kappa m)^{-2}$ and their difference for $\kappa \gtrsim \kappa^*$ is given by $m_f^{(3)}(0)-m_f^{(2)}(0) \approx 
\dfrac{4\sqrt{2}}{3\sqrt{3}}\left(\dfrac{\kappa-\kappa^*}{\kappa^*} \right)^{1/2}$.
\end{thebibliography}
\end{document}